\documentclass[preprint,amsmath,amssymb,11pt,english,aps,showkeys,showpacs]{revtex4}
\usepackage[english]{babel}
\usepackage{graphicx}
\usepackage{graphics}
\usepackage{amsmath}
\usepackage{dcolumn}
\usepackage{amssymb}
\usepackage{bm}
\usepackage{hyperref}

\begin{document}
\title{Relativistic quantum oscillators in the global monopole spacetime}

\author{E. A. F. Bragan\c{c}a}
\email{braganca@df.ufpe.br}
\affiliation{Departamento de F\'{i}sica, Universidade Federal de Pernambuco, 52171-900, Recife-PE, Brazil.}

\author{R. L. L. Vit\'oria}
\email{ricardo.vitoria@pq.cnpq.br}
\affiliation{Departamento de F\'isica e Qu\'imica, Universidade Federal do Esp\'irito Santo, Av. Fernando Ferrari, 514, Goiabeiras, 29060-900, Vit\'oria, ES, Brazil.}

\author{H. Belich}
\email{belichjr@gmail.com}
\affiliation{Departamento de F\'isica e Qu\'imica, Universidade Federal do Esp\'irito Santo, Av. Fernando Ferrari, 514, Goiabeiras, 29060-900, Vit\'oria, ES, Brazil.}

\author{E. R. Bezerra de Mello}
\email{emello@fisica.ufpb.br}
\affiliation{Departamento de F\'isica, Universidade Federal da Para\'i­ba, Caixa Postal 5008, Jo\~ao Pessoa-PB, 58051-900, Brazil.}

\begin{abstract}
We investigated the effects of the global monopole spacetime on the Dirac and Klein-Gordon relativistic quantum oscillators. In order to do this, we solve the Dirac and Klein-Gordon equations analytically and discuss the influence of this background, which is characterised by the curvature of the spacetime, on the energy profiles of these oscillators. In addition, we introduce a hard-wall potential and, for a particular case, determine the energy spectrum for relativistic quantum oscillators in this background.
\end{abstract}


\maketitle

\section{Introduction}
\label{sec1}
Grand unified theories predict that, in the early universe, as consequence of vacuum symmetry breaking phase transitions, topological defects could be produced \cite{kibble,VL} and the influence of these objects have been vastly investigated in various branches of physics \cite{dt, dt1, dt2, dt3, dt4, dt5, dt6, dt7}. In the context of condensed matter physics and gravitation, linear defects can be associated with the presence of torsion (dislocations) and curvature (disclinations) \cite{dt2, dt8}. Examples of well-known topological defects are domain wall \cite{dt9}, cosmic string \cite{dt10, dt11, dt12} and global monopole \cite{dt13}. In particular, the global monopole is a strong candidate to be observed in the field of observational cosmology \cite{dt} and it has been studied in several scenarios as $f(R)$ theory \cite{mg}, in the scalar self-energy for a charged particle \cite{mg1}, in nonrelativistic scattering \cite{mg2}, in the vacuum polarization for a massless spin-$\frac{1}{2}$ field \cite{mg3} and a massless scalar field \cite{mg4}. Also considering the presence of Wu-Yang magnetic monopole \cite{mg5} and in the gravitating magnetic monopole \cite{mg6}.

The effects of the global monopole also have been studied in the context of quantum mechanics.
Considering the nonrelativistic case, the Kratzer potential has been analyzed. In this case, the energy spectrum of the system is influenced by the defect \cite{mg7}, in a charged particle-magnetic monopole scattering \cite{mg7a} and on the harmonic oscillator \cite{mg7b}. For the relativistic case, there are studies on the hydrogen atom and pionic atom \cite{mg8}, on the exact solutions of scalar bosons in the presence of the Aharonov-Bohm and Coulomb potentials \cite{mg9}. In addition, the exact solutions of the Klein Gordon equation in the presence of a dyon, magnetic flux and scalar potential \cite{eug1} also have been analyzed in the nonrelativistic context. However, one point that has not been dealt with in the literature is the effects of the global monopole on the Dirac \cite{od} and Klein-Gordon \cite{okg} relativistic quantum oscillators. Therefore, in order to obtain solutions of bound states, in this paper, we deal with a fermionic field and a scalar field subject to the Dirac oscillator (DO) and the Klein-Gordon oscillator (KGO), respectively, in the global monopole spacetime. We also consider the presence of a hard-wall potential, that, for a particular case, the relativistic energy spectrum can be obtained.

Proposed by Monshisky and Szczepaniak, the relativistic quantum oscillator model for the spin-$\frac{1}{2}$ field, which it is known in the literature as DO \cite{od}, is derived from a nonminimum coupling in the Dirac equation, which preserves linearity in both the linear momentum and coordinates. In addition, at the non-relativistic limit, the modified Dirac equation is simplified in the Schr\"ondinger equation for the simple harmonic oscillator, however, added with a strong spin-orbit coupling term. It is noteworthy that the DO has been extensively studied, for example, in $(1+1)$-dimensions \cite{od1, od2, od3}, in $(2+1)$-dimensions \cite{od4, od5}, in the cosmic string spacetime \cite{od6, od6a,od6b}, in an Aharonov-Casher system \cite{od12}, in a rotating frame of reference \cite{od8}, in thermodynamic properties \cite{od8a, od8b} and in the context of gravity's rainbow \cite{od9}.

Inspired by the DO \cite{od}, Bruce and Minning have proposed a relativistic quantum oscillator model for the scalar field which it was known in the literature as the KGO \cite{okg} that, in the non-relativistic limit, is reduced to the oscillator described by the Schr\"ondinger equation \cite{okg1}. The KGO has been studied by a $\mathcal{PT}$-symmetric Hamiltonian \cite{okg2}, in noncommutative space \cite{okg3, okg4}, in spacetime with cosmic string \cite{okg5}, in a spacetime with torsion \cite{okg6, okg6a}, in a Kaluza-Klein theory \cite{okg7}, with noninertial effects \cite{okg8}, under effects of linear and Coulomb-type central potentials \cite{bf, me2, me3}, in thermodynamic properties \cite{pt, pt1} and in possible scenarios of Lorentz symmetry violation \cite{me1, me4}.

In this perspective, we intend to study the influence of topological defects, in particular the global monopole, in a high energy scenario (the early universe) which the DO and the KGO in the global monopole spacetime simulating this environment. The motivation for our proposal is to obtain clues of the oscillating behavior in Grand Unified Theories where we expect to occur vacuum symmetry breaking phase transition. A second motivation is that the results of our investigation can be applied to condensed matter systems.
Pointlike defects in spherical symmetry are present in elastic continuous solids as vacancies and impurities \cite{dt2}. In the high energy context, the solution of Einstein equations that describes the global monopole spacetime can be associated with these pointlike defects in solids \cite{tm3a}. This type of defect is defined by the distortion field of the medium and can be described by the Volterra process, where the vacancy in the continuous elastic medium can easily be imagined as follows: cut a sphere in half and remove its interior, and then reduce the sphere within a point. Impurities can be imagined in a reverse process, that is, to cut a sphere in the continuous elastic medium and remove its interior in order to add some extra matter into the medium to form the impurity.

The structure of this paper is organized as follows: in the Sec. (\ref{sec2}), we obtain the relativistic energy levels of a spin-$\frac{1}{2}$ fermionic field that interacts with the DO in the spacetime with a pointlike global monopole defect; in the Sec. (\ref{sec3}), we analyze a scalar field subject to the KGO in the spacetime with a pointlike global monopole defect and obtain the energy spectrum; in the Sec. (\ref{sec4}), we extend our initial discussions to the presence of a hard-wall confining potential; and in the Sec. \ref{sec5} we present our conclusions.

\section{DO in the global monopole spacetime}\label{sec2}

The global monopole spacetime is characterized by a line element that possesses a parameter related to the deficit angle  $\delta\Omega=8\pi^2G\eta_0^2$, with $\eta_0$ being the dimensionless volumetric mass density of the pointlike global monopole and $G$ is the gravitational Newton constant, and with the scalar curvature $R=R^{\mu}_{\mu}=2\frac{(1-\alpha^2)}{r^2}$. By working with the
units $c=\hbar=1$ and the signature $(-,+,+,+)$, the line element of the pointlike global monopole spacetime is written in the form \cite{mg8}:
\begin{eqnarray}\label{eq01}
ds^2=-dt^2+\frac{dr^2}{\alpha^2}+r^2d\theta^2+r^2\sin^2\theta d\varphi^2,
\end{eqnarray}
where $\alpha=1-8\pi^2G\eta_0^2<1$. In condensed mater, the line element above describes the effective metric produced in superfluid
$^3He-A$ by a monopole, with the angle deficit $\alpha>1$. In this case, the topological defect has a negative mass \cite{volovik1,volovik2}.

Henceforth, let us study the effects of the DO on the fermionic field in the global monopole spacetime. In curved spacetime, the Dirac equation is written in the form \cite{od6}
\begin{eqnarray}\label{eq02}
[i\gamma^{\mu}(x)(\partial_{\mu}+\Gamma_{\mu}(x))-m]\psi(x,t)=0,
\end{eqnarray}
where $\Gamma_{\mu}(x)$ is the spinorial connection component given in terms of the tetrad components and of the Christoffel symbols
\begin{eqnarray}\label{eq03}
\Gamma_{\mu}=\frac{1}{4}\gamma^{(a)}\gamma^{(b)}e^{\nu}_{ \ (a)}(\partial_{\mu}e_{\nu(b)}+\Gamma^{\lambda}_{\mu\nu}e_{\lambda(b)}).
\end{eqnarray}
The matrices $\gamma^{\mu}(x)$ are the generalized Dirac matrices for the background given by Eq. (\ref{eq01}) in terms of the standard Dirac matrices, that is, of the Minkowski spacetime Dirac matrices $\gamma^{(a)}$
\begin{eqnarray}\label{eq04}
\gamma^{\mu}(x)=e^{\mu}_{ \ (a)}(x)\gamma^{(a)},
\end{eqnarray}
which obey the relation of anticomutation $\gamma^{\mu}(x)\gamma^{\nu}(x)+\gamma^{\nu}(x)\gamma^{\mu}(x)=2g^{\mu\nu}(x)$.

The tetrad components $e^{\mu}_{ \ (a)}(x)$ obey the relation
\begin{eqnarray}\label{eq05}
g^{\mu\nu}(x)=e^{\mu}_{ \ (a)}(x)e^{\nu}_{ \ (b)}(x)\eta^{(a)(b)},
\end{eqnarray}
where $\mu,\nu=0,1,2,3$ are the indices corresponding to the curved spacetime, while $(a),(b)=0,1,2,3$ are indices corresponding to the Minkowski spacetime. We consider the following choice for the tetrad base in the pointlike global monopole spacetime (\ref{eq01}) \cite{mg8}:
\begin{eqnarray}\label{eq06}
e^{\mu}_{ \ (a)}(x)=\left[
\begin{array}{cccc}
 1 & 0 & 0 & 0 \\
 0 & \alpha\sin\theta\cos\varphi & \frac{\cos\theta\cos\varphi}{r} & -\frac{\sin\varphi}{r\sin\theta} \\
 0 & \alpha\sin\theta\sin\varphi & \frac{\cos\theta\sin\varphi}{r} & \frac{\cos\varphi}{r\sin\theta} \\
 0 & \alpha\cos\theta & -\frac{\sin\theta}{r} & 0 \\
\end{array}
\right].
\end{eqnarray}
In this representation the matrices $\gamma^{\mu}(x)$ obey the relations
\begin{eqnarray}\label{eq07}
\gamma^{0}=\gamma^{t}; \ \ \ \gamma^{1}=\alpha\vec{\gamma}.\hat{r}=\alpha\gamma^{(r)}; \ \ \ \gamma^{2}=\frac{1}{r}\vec{\gamma}.\hat{\theta}=\frac{1}{r}\gamma^{(\theta)}; \ \ \ \gamma^{3}=\frac{\vec{\gamma}.\hat{\varphi}}{r\sin\theta}=\frac{\gamma^{(\varphi)}}{r\sin\theta},
\end{eqnarray}
where $\hat{r}$, $\hat{\theta}$ and $\hat{\varphi}$ are the spherical versors. In addition, the nonzero spinor connections are
\begin{eqnarray}\label{eq08}
\Gamma_2&=&\frac{(\alpha-1)}{2}[\gamma^{(1)}\gamma^{(3)}\cos\varphi+\gamma^{(2)}\gamma^{(3)}\sin\varphi]; \nonumber \\
\Gamma_3&=&-\frac{(\alpha-1)}{2}[\gamma^{(1)}\gamma^{(2)}\sin\theta+\gamma^{(1)}\gamma^{(3)}\cos\theta\sin\varphi-\gamma^{(2)}\gamma^{(3)}\cot\theta\cos\varphi]\sin\theta,
\end{eqnarray}
where
\begin{eqnarray}\label{eq09}
\gamma^{2}\Gamma_{2}+\gamma^{3}\Gamma_{3}=\frac{(\alpha-1)}{r}\vec{\gamma}.\hat{r}.
\end{eqnarray}

Given all this, we can now investigate the effects of the DO on a fermionic field in the pointlike global monopole spacetime. To introduce the DO in the Dirac equation (\ref{eq02}) we must use the following coupling \cite{od, od6}: $\partial_{r}\rightarrow\partial_{r}+m\omega\beta r$, where $\omega$ is the angular frequency of the DO. Then, the Dirac equation (\ref{eq02}) for the spacetime under consideration given by the Eq. (\ref{eq01}), with the Eqs. (\ref{eq06}), (\ref{eq07}), (\ref{eq08}) and (\ref{eq09}), becomes
\begin{eqnarray}\label{eq10}
\left[i\gamma^{(t)}\partial_t+i\gamma^{(r)}\left(\alpha\partial_r+\frac{(\alpha-1)}{r}+m\omega\alpha\beta r\right)+i\frac{\gamma^{(\theta)}}{r}\partial_{\theta}+i\frac{\gamma^{(\varphi)}}{r\sin\theta}\partial_{\varphi}-m\right]\psi=0,
\end{eqnarray}
with
\begin{eqnarray}\label{eq11}
\gamma^{(t)}=\left[
\begin{array}{cc}
 I_2 & 0 \\
 0 & -I_2 \\
\end{array}
\right]; \ \ \
\gamma^{(i)}=\left[
\begin{array}{cc}
 0 & \sigma^{i} \\
 -\sigma^{i} & 0 \\
\end{array}
\right],
\end{eqnarray}
where $I_2$ is order two identity matrix and $\sigma^{i}$ are Pauli matrices.

The complete set of solutions for Eq. (\ref{eq10}) is \cite{bjo, greiner}
\begin{eqnarray}\label{eq12}
\psi(\vec{r},t)=\frac{1}{r}\left[
\begin{array}{c}
  if(r)\Phi_{j,m_l}(\theta,\varphi) \\
    g(r)(\vec{\sigma}.\hat{r})\Phi_{j,m_l}(\theta,\varphi) \\
\end{array}
\right]e^{-i\mathcal{E}t},
\end{eqnarray}
where $\Phi_{j,m_l}$ are the spinor spherical harmonics, $f(r)$ and $g(r)$ are radial wave spinor functions and $\mathcal{E}$ is the energy of the system. Note that the Eq. (\ref{eq11}) presents a well defined parity under the transformation $\vec{r}\rightarrow\vec{r'}=-\vec{r}$ \cite{bjo}. Hence, by substituting the Eq. (\ref{eq12}) into Eq. (\ref{eq10}), we obtain radial differential equations
\begin{subequations}
\begin{equation}\label{eq13a}
(\mathcal{E}-m)f+\alpha\frac{dg}{dr}-\frac{\kappa}{r}g-m\omega\alpha rg=0;
\end{equation}
\begin{equation}\label{eq13b}
(\mathcal{E}+m)g-\alpha\frac{df}{dr}-\frac{\kappa}{r}f-m\omega\alpha rf=0,
\end{equation}
\end{subequations}
where we use the definitions
\begin{eqnarray}\label{eq14}
(\vec{\sigma}.\vec{A})(\vec{\sigma}.\vec{B})=\vec{A}.\vec{B}+i\vec{\sigma}.(\vec{A}\times\vec{B}); \ \vec{\sigma}.\vec{L}\Phi_{j,m_l}=-(1+\kappa)\Phi_{j,m_l}; \ \vec{L}=-i\left(\hat{\varphi}\frac{\partial}{\partial\theta}-\frac{\hat{\theta}}{\sin\theta}\frac{\partial}{\partial\varphi}\right),
\end{eqnarray}
with $\kappa=\mp j+\frac{1}{2}$, $j=l\pm\frac{1}{2}$ and $l=0,1,2,\ldots$.

Multiplying \eqref{eq13a} by $(\mathcal{E}+m)$ and using \eqref{eq13b} we obtain
\begin{eqnarray}\label{eq15}
\frac{d^2f}{dr^2}-\frac{\kappa(\kappa+\alpha)}{\alpha^2r^2}f-m^2\omega^2r^2f+\frac{[\mathcal{E}^2-m^2+m\omega\alpha(\alpha-2\kappa)]}{\alpha^2}f=0.
\end{eqnarray}
With the purpose of solving this radial equation, let us write
\begin{eqnarray}\label{eq16}
f(r)=\frac{F(r)}{\sqrt{r}},
\end{eqnarray}
then, we obtain the following equation for the function $F(r)$:
\begin{eqnarray}\label{eq17}
\frac{d^2F}{dr^2}-\frac{1}{r}\frac{dF}{dr}+\frac{3}{4r^2}F-\frac{\kappa(\kappa+\alpha)}{\alpha^2r^2}F-m^2\omega^2r^2F
+\frac{[\mathcal{E}^2-m^2+m\omega\alpha(\alpha-2\kappa)]}{\alpha^2}F=0.
\end{eqnarray}
We proceed with a change of variables given by $s=m\omega r^2$, and thus we obtain:
\begin{eqnarray}\label{eq18}
\frac{d^2F}{ds^2}+\frac{(1-4\mu^2)}{4s^2}F+\frac{\nu}{s}F-\frac{1}{4}F=0,
\end{eqnarray}
where we define new parameters given by
\begin{eqnarray}\label{eq19}
\mu=\frac{\sqrt{\alpha^2+4\kappa(\kappa+\alpha)}}{4\alpha}; \ \ \ \nu=\frac{\mathcal{E}^2-m^2+m\omega\alpha(\alpha-2\kappa)}{4m\omega\alpha^2}.
\end{eqnarray}

Hence, the Eq. (\ref{eq18}) is the well-known Whittaker differential equation \cite{kdm} and $F(s)$ is the Whittaker function which can be written in terms of confluent hypergeometric function$ \ _{1}F_1(s)$ \cite{kdm, arf}
\begin{eqnarray}\label{eq20}
F(s)=s^{\frac{1}{2}+\mu}e^{-\frac{1}{2}s} \ _{1}F_{1}\left(\mu-\nu+\frac{1}{2},2\mu+1;s\right).
\end{eqnarray}
It is well known that the confluent hypergeometric series becomes a polynomial of degree $n$ by imposing that $\mu-\nu+\frac{1}{2}=-n$, where $n=0,1,2,3\ldots$. With this condition, we obtain
\begin{eqnarray}\label{eq21}
\mathcal{E}_{l,n}=\pm\sqrt{m^2+4m\omega\alpha^2\left(n+\frac{\sqrt{\alpha^2+4\kappa(\kappa+\alpha)}+2\kappa}{4\alpha}+\frac{1}{4}\right)}.
\end{eqnarray}

\begin{figure}[h]
	\centering
	{\includegraphics[width=0.48\textwidth]{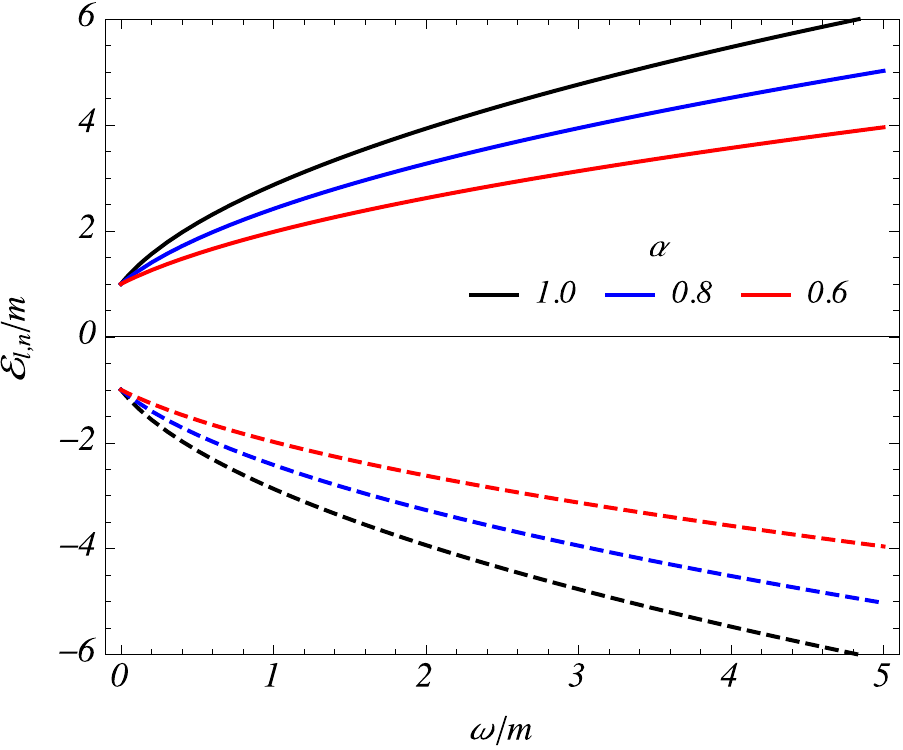}}
	\hfill
	{\includegraphics[width=0.48\textwidth]{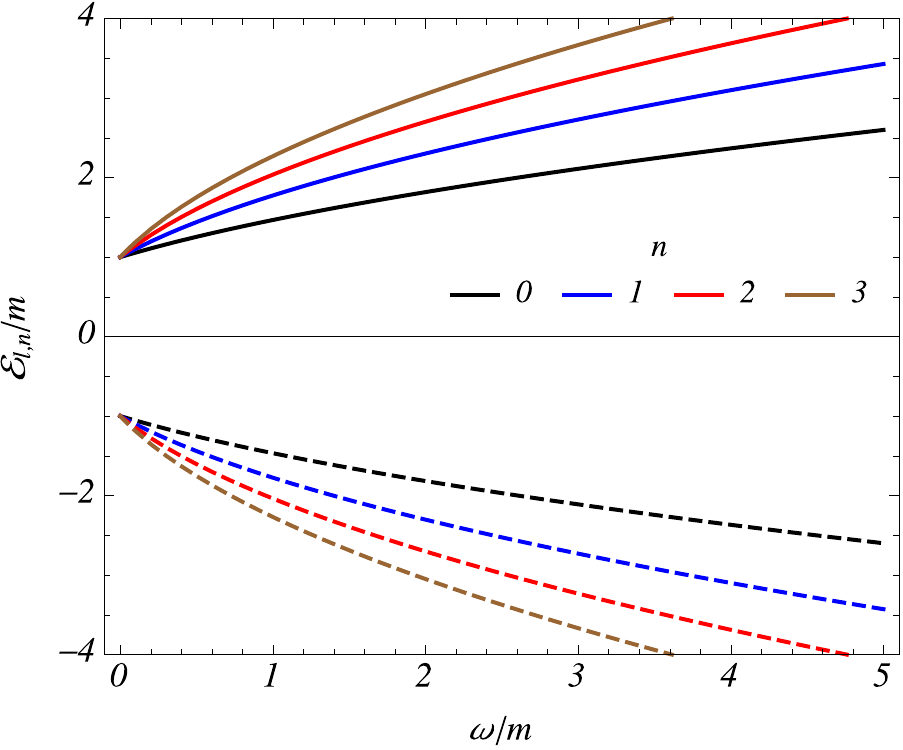}}
	\caption{Energy profile as function of $\omega/m$ for $n=1$ (left plot) and $\alpha=0.5$ (right plot). The full lines are the positive energy while
	the dashed lines are for the negative energy. In both plots we consider $\kappa=1/2$.}
	\label{fig01}
\end{figure}

The Eq. (\ref{eq21}) gives us the relativistic energy spectrum of a spin-$\frac{1}{2}$ fermionic field subject to the DO in a pointlike global monopole spacetime. We can note that the spacetime topology influences the relativistic energy levels of the system through the presence of the parameter associated with the topological defect responsible by the curvature of the spacetime, that is, $\alpha$. By making $\alpha\rightarrow1$, we recover the energy spectrum of a spin-$\frac{1}{2}$ fermionic field subject to the DO in $(3+1)$-dimensions in the Minkowski spacetime.
In Fig. \ref{fig01} we have plotted the profile of the energy \eqref{eq21} as function of $\omega/m$. In the left plot, we consider different
values of the parameter $\alpha$ and $n=1$, with $\alpha=1$ meaning the absence of the global monopole.
In the right plot, we consider different energy levels for a fixed value of the parameter which characterises the presence of the
global monopole. From the figure, we show how $|\mathcal{E}_{l,n}|$ is diminished by the presence of the topological defect and increases
with $n$. Besides that, we note the as $\omega\rightarrow 0$, $\mathcal{E}_{l,n}\rightarrow \pm m$, for any value of $n$.

The set of eigenfunctions are given by
\begin{eqnarray}\label{eq22}
f_{l,n}(s)&=&C_{l,n}\,\left(m\omega\right)^{\frac{1}{4}}s^{\mu+\frac{1}{4}}e^{-\frac{1}{2}s} \ _{1}F_{1}\left(-n, 2\mu+1;s\right),
\label{Fdirac}
\end{eqnarray}
and
\begin{eqnarray}
g_{l,n}(s)&=&\frac{\sqrt{m\omega}}{(\mathcal{E}_{l,n}+m)}\left(2\alpha\sqrt{s}\frac{d}{ds}+\frac{\kappa}{\sqrt{s}}+\alpha\sqrt{s}\right)f_{l,n}(s).
\end{eqnarray}
The above solution can be obtained by using \eqref{Fdirac}, which gives
\begin{eqnarray}
g_{l,n}(s)&=&C_{l,n} \,\frac{(m\omega)^{3/4}e^{-s/2}s^{\mu-1/4}}{2(\mathcal{E}_{l,n}+m)}
\left[(\alpha+2\kappa+4\alpha\mu)_1F_1\left(-n,2\mu+1;s\right)\right.\nonumber\\
&&\left.-\frac{4ns\alpha}{(2\mu+1)} \,_1F_1\left(-n+1,2\mu+2;s\right)\right].
\label{Gdirac}
\end{eqnarray}

In the Eqs. \eqref{Fdirac} and \eqref{Gdirac}, $C_{l,n}$ is a constant which can be determined by the normalization
condition for the radial eigenfunctions
\begin{equation}
\frac{1}{\alpha}\int_0^\infty dr \left[|f_{l,n}(r)|^2+|g_{l,n}(r)|^2\right]=1.
\label{normRDO}
\end{equation}
To solve the integrals of the radial eigenfunctions we write the confluent hypergeometric functions in terms of the
associated Laguerre polynomials by the relation \cite{arf}
\begin{equation}
_1F_1(-n,\gamma+1,x)=\frac{n!\gamma!}{(n+\gamma)!}
L_n^\gamma(x).
\label{HyperToLaguerre}
\end{equation}
Then, taking into account that $s=m\omega r^2$ and with the help of \cite{prudnikov} to solve the integrals, the normalization
constant is given by
\begin{eqnarray}
C_{l,n}&=&\sqrt{\frac{2\alpha(n+2\mu)!}{(2\mu)!}}\left\{n!(2\mu)!+\frac{m\omega}{4(\mathcal{E}+m)^2}
\left[n!(\alpha+2\kappa+4\alpha\mu)^2(2\mu-1)!\right.\right.\nonumber\\
&&\left.\left.-8\alpha n!(\alpha+2\kappa+4\alpha\mu)(2\mu)!\delta_{n,n\geq 1}
+16n\alpha^2 n!(2\mu)!(2\mu+1)^2\right]\right\}^{-1/2}.
\nonumber\\
\label{DiracConstant}
\end{eqnarray}
The presence of $\delta_{n,n\geq 1}$ in \eqref{DiracConstant} means that 
for the fundamental state, $n=0$, the corresponding term vanishes. This is due to the fact
that this term is obtained from the derivative of the confluent hypergeometric function in
the solution $f_{l,n}(s)$ which is zero for the fundamental state$^{\footnotemark[1]}$.
\footnotetext[1]{In fact, there is a $\delta_{n,n\geq 1}$ following the second confluent hypergeometric function in \eqref{Gdirac} and also in the last term inside the brackets in \eqref{DiracConstant}. But we decided to omit it because the presence of the quantum number $n$ in both terms, 
thus avoiding being redundant.}

\begin{figure}[h]
	\centering
	{\includegraphics[width=0.48\textwidth]{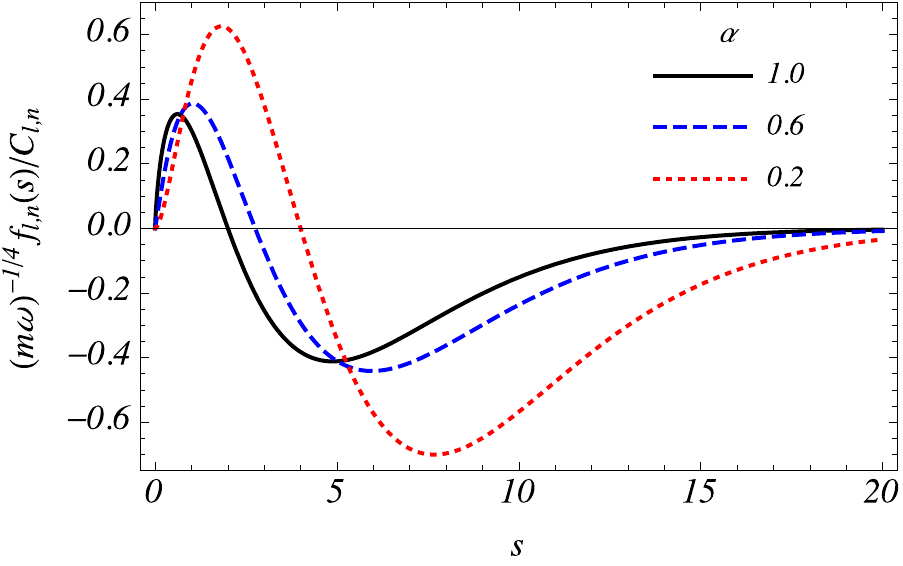}}
	\hfill
	{\includegraphics[width=0.48\textwidth]{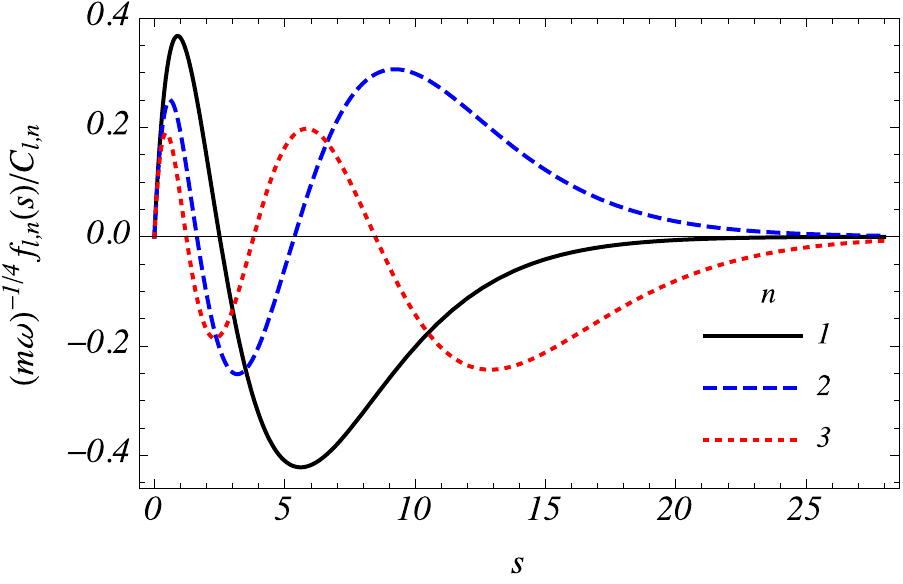}}
	\caption{Profile of the eigenfunction $f_{l,n}(s)$ as function of the coordinate $s$. In the left plot, we consider different values
	of the parameter $\alpha$ and $n=1$. In the right plot, we have $\alpha=0.5$, and different energy levels. In both plots we consider $\kappa=1/2$.}
	\label{functionsDO}
\end{figure}

From Eqs. \eqref{Fdirac} and \eqref{Gdirac} we can see that the these eigenfunctions are influenced by the spacetime topology through the presence of the parameter associated with the topological defect. In the Fig. \ref{functionsDO} we have plotted the behavior of the eigenfunction $f_{l,n}(s)$ as function of the coordinate
$s$. In the left plot, we consider a fixed energy level and different values of the parameter $\alpha$, while in the second one we consider
a fixed value of $\alpha$ and different energy levels. From the figure we note that for small $s$ the eigenfunction is highly influenced by the
presence of the global monopole. This eigenfunction decreases with the coordinate $s$ and far away from the defect ($s\rightarrow \infty$) becomes negligible.
In addition, in Table \ref{Tab1} show the numerical values of the normalization constant \eqref{DiracConstant} for different values of $\alpha$ and different
energy levels, considering a fixed value of $l$, $\kappa$, and $\omega/m$.

\begin{table}[h]
\begin{tabular}{|c|c|c|c|}
\hline
$ \ C_{l,n} \ $ & \ $\alpha=1$    \  & \ $\alpha=0.8$  \  & \ $\alpha=0.6$ \   \\
\hline
$C_{0,0}$ & 1.1053 & 0.9132 & 0.6669 \\
\hline
$C_{0,1}$ & 1.7491 & 1.3963 & 0.9523 \\
\hline
$C_{0,2}$ & 2.0035 & 1.5911 & 1.0702 \\
\hline
\end{tabular}
\caption{Numerical values of the normalization constant, Eq. \eqref{DiracConstant}, considering different values of $\alpha$ for $l=0$, $\kappa=1$ and
$\omega/m=0.5.$}
\label{Tab1}
\end{table}

\section{KGO in the global monopole spacetime}\label{sec3}

In this section, we investigate the scalar field under effects of the KGO in the pointlike global monopole spacetime. The KGO is described by introducing a coupling into the Klein-€"Gordon equation as $\hat{p}_{\mu}\rightarrow\hat{p}_{\mu}+im\omega X_{\mu}$, where $m$ is the rest mass of the scalar field,  $\omega$ is the angular frequency of the KGO and $X_{\mu}=(0,r,0,0)$ \cite{okg6, okg6a, okg7}. The Klein-Gordon equation in the spacetime with a pointlike global monopole (\ref{eq01}) can be written as \cite{okg7}
\begin{eqnarray}\label{eq23}
\frac{1}{\sqrt{-g}}(\partial_{\mu}+m\omega X_{\mu})(\sqrt{-g}g^{\mu\nu})(\partial_{\nu}-m\omega X_{\nu})\phi-\xi R\phi-m^2\phi=0,
\end{eqnarray}
where $g=\text{det}(g_{\mu\nu})$, $g^{\mu\nu}$ is inverse metric tensor, $\xi$ is an arbitrary coupling constant and $R$ the curvature scalar already reported in the previous section. In this way, from the Eq. (\ref{eq01}), the Klein-Gordon equation (\ref{eq23}) becomes
\begin{eqnarray}\label{eq24}
-\partial_t^2\phi&+&\frac{\alpha^2}{r^2}(\partial_r+m\omega r)(r^2\partial_r\phi-m\omega r^3\phi)+\frac{1}{r^2\sin\theta}\partial_\theta(\sin\theta\partial_\theta\phi)+\frac{1}{r^2\sin^2\theta}\partial_\varphi^2\phi \nonumber \\
&-&\frac{2\xi(1-\alpha^2)}{r^2}\phi-m^2\phi=0.
\end{eqnarray}

The solution to the Eq. (\ref{eq24}) can be written in the form
\begin{eqnarray}\label{eq25}
\phi(r,\theta,\varphi,t)=e^{-i\mathcal{E}t}R(r)Y_{l,m_l}(\theta,\varphi),
\end{eqnarray}
where $Y_{l,m_l}(\theta,\varphi)$ are the spherical harmonics and $f(r)$ is a radial wave function. Then, by substituting the solution (\ref{eq25}) into the Eq. (\ref{eq24}), we have
\begin{eqnarray}\label{eq26}
\frac{\alpha^2}{r^2R}\left(\frac{d}{dr}+m\omega r\right)\left(r^2\frac{dR}{dr}-m\omega r^3R\right)-\frac{2\xi}{r^2}(1-\alpha^2)+(\mathcal{E}^2-m^2) \nonumber \\
-\frac{1}{r^2}\left[-\frac{1}{\sin\theta Y_{l,m_l}}\frac{\partial}{\partial\theta}\left(\sin\theta\frac{\partial Y_{l,m_l}}{\partial\theta}\right)
-\frac{1}{\sin^2\theta Y_{l,m_l}}\frac{\partial^2Y_{l,m_l}}{\partial\varphi^2}\right]=0,
\end{eqnarray}
or
\begin{eqnarray}\label{eq27}
\frac{d^2R}{dr^2}+\frac{2}{r}\frac{dR}{dr}-\frac{[2\xi(1-\alpha^2)+l(l+1)]}{\alpha^2r^2}R-m^2\omega^2r^2R
+\frac{(\mathcal{E}^2-m^2-3m\omega\alpha^2)}{\alpha^2}R=0,
\end{eqnarray}
where we use the definition \cite{greiner}
\begin{eqnarray}\label{eq28}
\left[\frac{1}{\sin\theta}\frac{\partial}{\partial\theta}\left(\sin\theta\frac{\partial}{\partial\theta}\right)+
\frac{1}{\sin^2\theta}\frac{\partial^2}{\partial\varphi^2}\right]Y_{l,m_l}(\theta,\varphi)=-l(l+1)Y_{l,m_l}(\theta,\varphi).
\end{eqnarray}

With the purpose of solving the radial differential equation (\ref{eq27}), let us write
\begin{eqnarray}\label{eq29}
R(r)=\frac{\bar{W}(r)}{r^{\frac{3}{2}}},
\end{eqnarray}
then, we obtain the following equation for the function $W(r)$:
\begin{eqnarray}\label{eq30}
\frac{d^2\bar{W}}{dr^2}-\frac{1}{r}\frac{d\bar{W}}{dr}+\frac{3\bar{W}}{4r^2}-\frac{[2\xi(1-\alpha^2)+l(l+1)]}{\alpha^2r^2}\bar{W}-m^2\omega^2r^2\bar{W}
+\frac{(\mathcal{E}^2-m^2-3m\omega\alpha^2)}{\alpha^2}\bar{W}=0.
\end{eqnarray}

Let us define $s=m\omega r^2$, then the Eq. (\ref{eq30}) becomes
\begin{eqnarray}\label{eq31}
\frac{d^2\bar{W}}{ds^2}+\frac{(1-4\bar{\mu}^2)}{4s^2}\bar{W}+\frac{\bar{\nu}}{s}\bar{W}-\frac{1}{4}\bar{W}=0,
\end{eqnarray}
where we have defined the parameters
\begin{eqnarray}\label{eq32}
\bar{\mu}=\frac{\sqrt{\alpha^2+8\xi(1-\alpha^2)+4l(l+1)}}{4\alpha}; \ \ \ \bar{\nu}=\frac{\mathcal{E}^2-m^2-3m\omega\alpha^2}{4m\omega\alpha^2}.
\end{eqnarray}
Note that the Eq. (\ref{eq31}) is also a Whittaker differential equation \cite{kdm}. Then, by following the steps from Eqs. (\ref{eq20}) to (\ref{eq21}), we obtain
\begin{eqnarray}\label{eq33}
\mathcal{E}_{l,n}=\pm\sqrt{m^2+4m\omega\alpha^2\left(n+\frac{\sqrt{\alpha^2+8\xi(1-\alpha^2)+4l(l+1)}}{4\alpha}+\frac{5}{4}\right)}.
\end{eqnarray}

\begin{figure}[h]
	\centering
	{\includegraphics[width=0.48\textwidth]{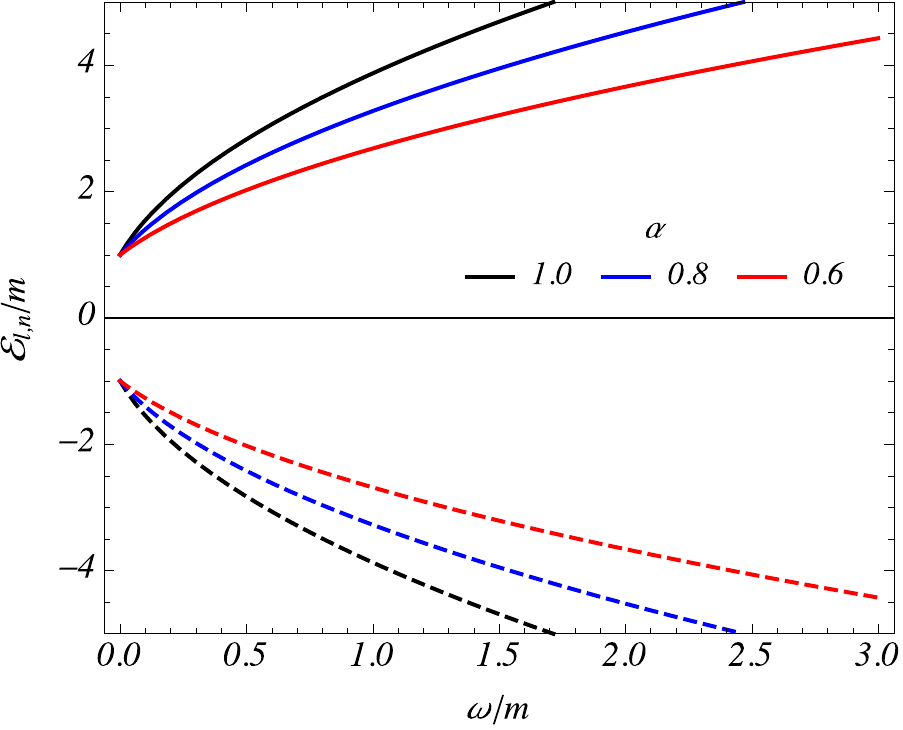}}
	\hfill
	{\includegraphics[width=0.48\textwidth]{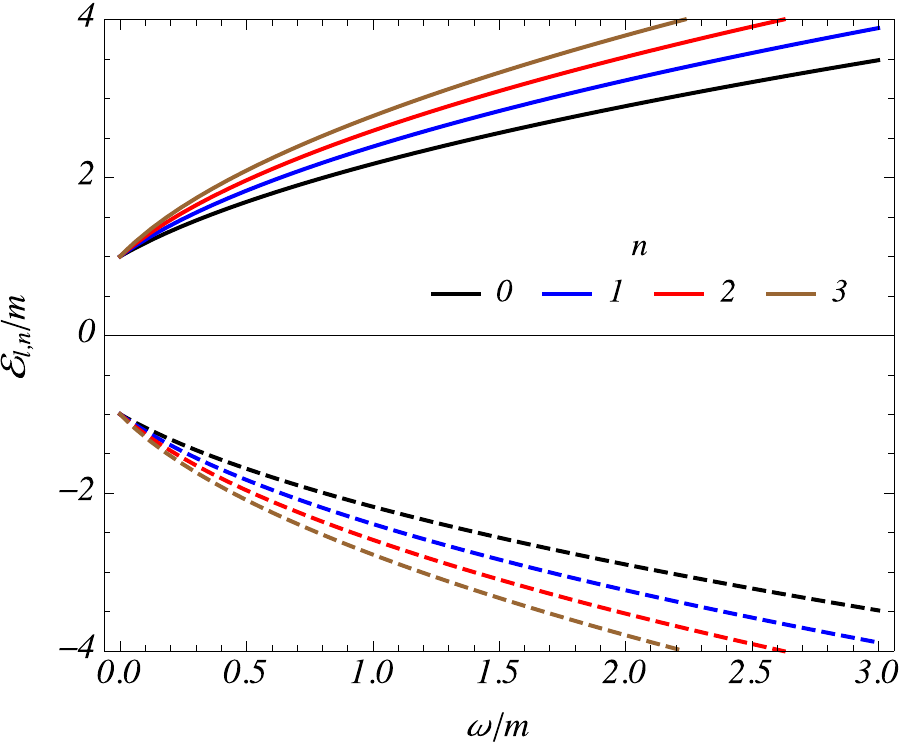}}
	\caption{Energy profile of the KGO as function of $\omega/m$. In the left plot we have different values of $\alpha$ and $n=1$,
	while in the right plot we consider $\alpha=0.5$ and different energy levels. Also $l=0$ and $\xi=0$ in both plots. The full and dashed curves
	represent the positive and negative energies, respectively.}
	\label{fig02}
\end{figure}

Then, the Eq. (\ref{eq33}) is the relativistic energy spectrum that stems from the interaction of the scalar field with the KGO in the spacetime with a pointlike global monopole. As in the case of the DO, the presence of the parameter $\alpha$ into the Eq. (\ref{eq33}), modifies the relativistic energy levels of the system. By making $\alpha\rightarrow1$, we recover the relativistic energy levels of a scalar field subject to the KGO in $(3+1)$-dimensions in the Minkowski spacetime.
In the Fig. \ref{fig02} we have plotted the energy profiles of the KGO as function of $\omega/m$. In the left plot,
we consider for different values of the parameter which characterises the presence of the topological defect, $\alpha$. Again,
the value $\alpha=1$ means that the global monopole is absent.
In the right plot, we keep fixed the value of $\alpha$ and we consider different energy levels. Similarly to the DO case,
we note that the $|\mathcal{E}_{l,n}|$ is diminished by the presence of the global monopole. In addition, the figure shows how
the energy of the oscillator increase with $n$ and for small values of the $\omega/m$, the energy of the KGO goes to $\pm m$.

The radial eigenfunctions, then, are given by
\begin{eqnarray}\label{eq34}
R_{l,n}(s)=D_{l,n}(m\omega)^{\frac{3}{4}}s^{\bar{\mu}-\frac{1}{4}}e^{-\frac{1}{2}s} \ _{1}F_{1}\left(-n,2\bar{\mu}+1;s\right),
\end{eqnarray}
that is, the system eigenfunctions are also influenced by the topology of the spacetime through the presence of the parameter associated with the topological defect, $\alpha$, in the structure of the Eq. (\ref{eq34}).
In the above expression, $D_{l,n}$ is a constant which can be determined by the normalization condition for
the radial eigenfunction
\begin{equation}
\frac{1}{\alpha}\int_0^\infty dr r^2|R(r)|^2=1.
\end{equation}
Following a similar procedure done to the DO, the normalization constant is given by
\begin{equation}
D_{l,n}=\frac{1}{(2\bar{\mu})!}\sqrt{\frac{2\alpha(n+2\bar{\mu})!}{n!}}.
\label{KGconstant}
\end{equation}

\begin{figure}[h]
	\centering
	{\includegraphics[width=0.48\textwidth]{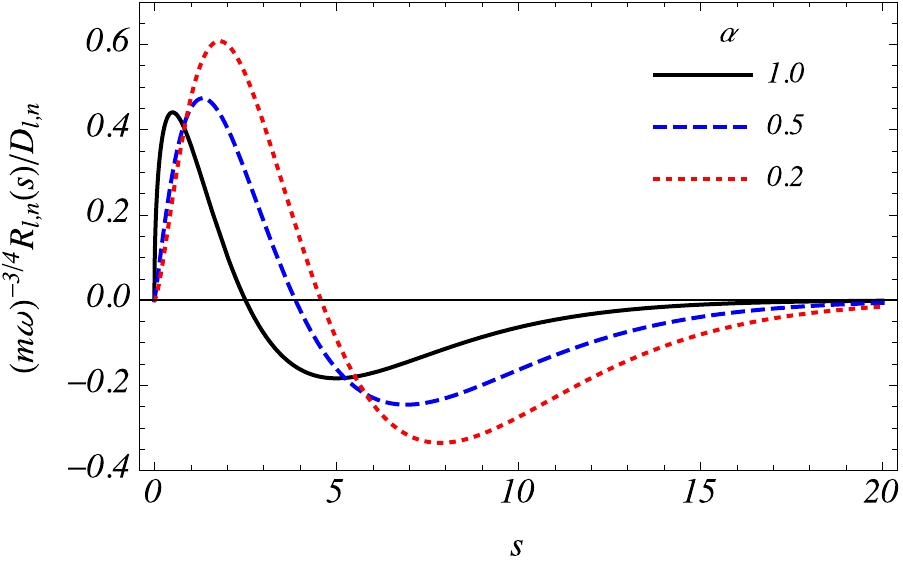}}
	\hfill
	{\includegraphics[width=0.48\textwidth]{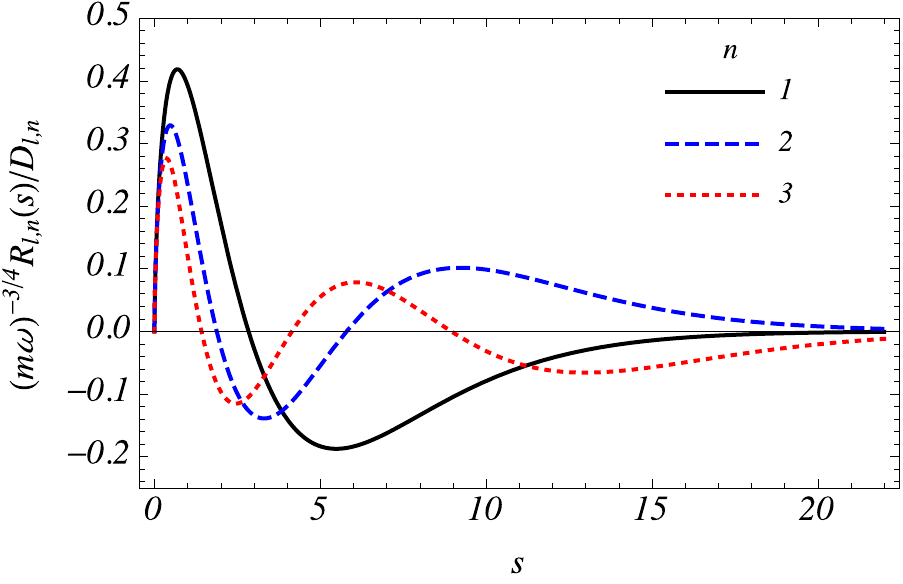}}
	\caption{Profile of the eigenfunction $R_{l,n}(s)$ as function of the coordinate $s$. In the left plot, we have $n=1$ and different values
	of the parameter $\alpha$. In the right plot, we have $\alpha=0.8$ and different energy levels. In addition, we have $l=1$ and $\xi=0$ in both plots. }
	\label{functionKGO}
\end{figure}

In Fig. \ref{functionKGO} we have plotted the eigenfunction $R_{l,n}(s)$ as function of the
coordinate $s$. In the left plot, we consider a fixed energy level and different values of the parameter $\alpha$, while in the right one
we consider a fixed value of $\alpha$ and different energy levels. Similarly to the DO case, the eigenfunction is highly influenced by the presence of the
global monopole in regions near to the defect. When we consider regions far away from the global monopole ($s\rightarrow \infty$)
the eigenfunction decreases and the
effect of the defect becomes negligible. In addition, in Table \ref{Tab2} we show numerical values of the normalization constant
above, considering different energy levels and different values of the parameter $\alpha$, for a fixed value of $\xi$ and the quantum number $l$.

\begin{table}[h]
\begin{tabular}{|c|c|c|c|}
\hline
$ \ D_{l,n} \ $ & \ $\alpha=1$    \  & \ $\alpha=0.8$  \  & \ $\alpha=0.6$ \   \\
\hline
$D_{0,1}$ & 1.5022 & 1.3436 & 1.1636 \\
\hline
$D_{0,2}$ & 1.8398 & 1.6456 & 1.4251 \\
\hline
$D_{0,3}$ & 2.0570 & 1.8398 & 1.59338 \\
\hline
\end{tabular}
\caption{Numerical values of the normalization constant, Eq. \eqref{KGconstant}, considering different values of $\alpha$ for $l=\xi=0$.}
\label{Tab2}
\end{table}

\section{Effects of a hard-wall potential}
\label{sec4}

Now, in this section, we want to restrict the motion of the relativistic spin-$\frac{1}{2}$ fermionic and scalar fields to a region where a hard-wall confining potential is present. This type of confinement is important because it is a very good approximation to consider when discussing the quantum properties of a gas molecule system and other particles, which are necessarily confined in a box of certain dimensions. The hard-wall potential has been studied in rotating effects on the scalar field \cite{dt7}, in the KGO under effects of linear topological defects \cite{okg6, okg8}, in noninertial effects on a nonrelativistic Dirac particle \cite{kb}, in a Landau-Aharonov-Casher system \cite{kb1}, on a Dirac neutral particle in analogous way to a quantum dot \cite{kb2}, on the harmonic oscillator in an elastic medium with a spiral dislocation \cite{kb3}, on persistent currents for a moving neutral particle with no permanent electric dipole moment \cite{kb4} and in a Landau-type quantization from a Lorentz symmetry violation background with crossed electric and magnetic fields \cite{kb5}.

\subsection{DO}\label{sec4a}
Here we confine the spin-$\frac{1}{2}$ fermionic field subject to the DO in the global monopole spacetime. In order
to do that, we impose that the fermionic field obeys the MIT bag boundary condition at a finite radius $s_0$:
\begin{equation}
(1+in_\mu \gamma^\mu)\psi|_{s=s_0}=0,
\label{MIT}
\end{equation}
with $n_\mu$ is the outward oriented normal (with respect to the region under consideration) to the boundary. As we shall consider
the region inside the bag, we have that $n_\mu=-\frac{\delta_\mu ^1}{\alpha}$.
Taking into consideration \eqref{eq12}, the Eq. \eqref{MIT} becomes
\begin{equation}
f_{n,l}(s_0)-g_{n,l}(s_0)=0,
\end{equation}
and by using \eqref{eq22} one finds
\begin{eqnarray}
&&\frac{2\alpha\sqrt{m\omega s_0}}{\mathcal{E}+m}\frac{(\mu-\nu-1/2)}{2\mu+1} \ _1F_1\left(\mu-\nu+\frac{3}{2},2\mu+2,s_0\right)+\nonumber\\
&&\left\{\frac{\sqrt{m\omega}}{\mathcal{E}+m}\left[\frac{2\alpha}{s_0^{1/2}}(\mu+1/4)+\frac{\kappa}{s_0^{1/2}} \right]-1\right\} \
_1F_1\left(\mu-\nu+\frac{1}{2}, 2\mu+1,s_0\right)=0
\label{MITeq1}
\end{eqnarray}

\begin{figure}[h]
	\centering
	{\includegraphics[width=0.48\textwidth]{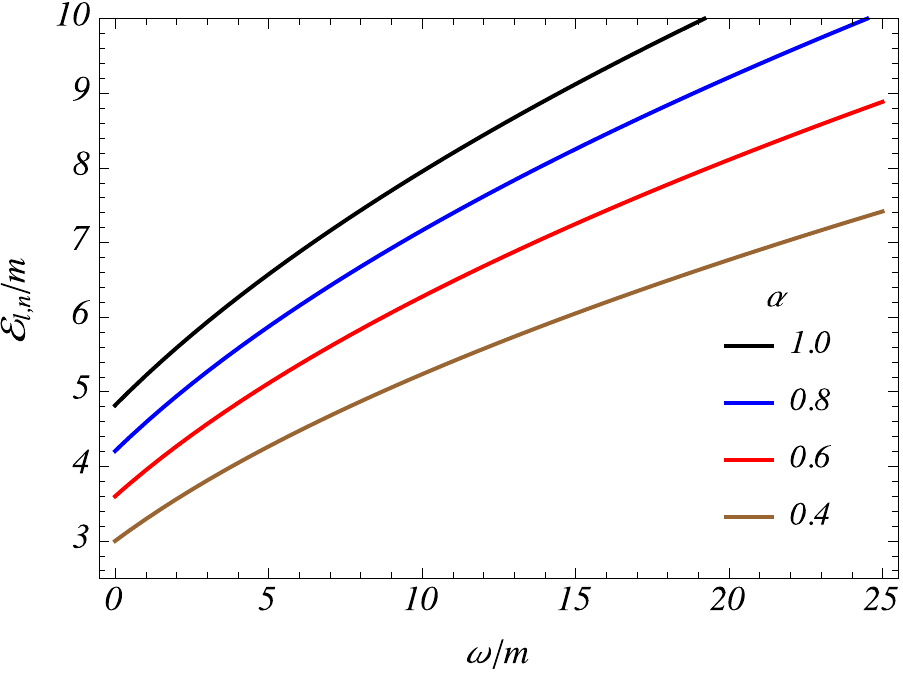}}
	\hfill
	{\includegraphics[width=0.48\textwidth]{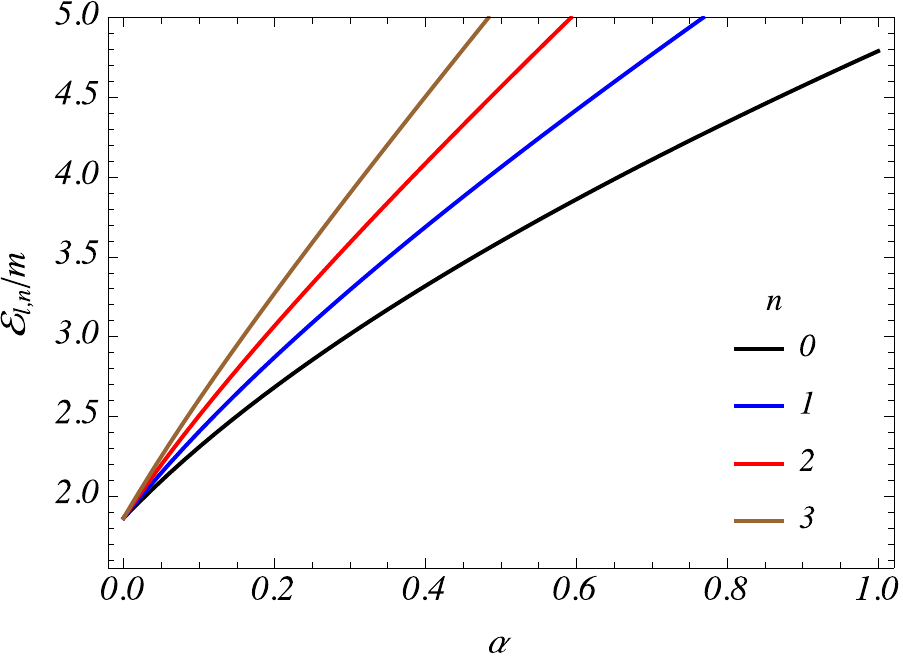}}
	\caption{Positive energy profile of the DO subject to the MIT bag boundary condition as function of $\omega/m$ (left plot)
	and $\alpha$ (right plot). In the left plot we have different values of $\alpha$ and $n=1$,
	while in the right plot we consider $\omega/m=2.5$ and different energy levels. In both plots we have $mr_0=2.5$ and $\kappa=5/2$.}
	\label{fig03}
\end{figure}

Let us consider $\nu\gg1$. In this limit we can use the formula \cite{abr}
\begin{equation}
\lim_{a\rightarrow \infty} \ _1F_1(a,b;z)=\Gamma(b)\left[z\left(\frac{b}{2}-a\right)\right]^{1/4-b/2}\pi^{-1/2} \ e^{z/2}
\cos\left[\sqrt{z\left(2b-4a\right)}-\frac{b}{2}\pi+\frac{\pi}{4}\right],
\end{equation}
for the parameters $b$ and $z$ fixed. Taking into account the previous equation, we have for the confluent hypergeomentric
functions
\begin{equation}
\lim_{\nu \rightarrow \infty} \ _1F_1\left(\mu-\nu+\frac{3}{2},2\mu+2,s_0\right)\approx
\frac{\Gamma(2\mu+2)}{\sqrt{\pi}}\frac{\cos[2\sqrt{s_0(\nu-1/2)}-\mu\pi-3\pi/4]}{[s_0(\nu-1/2)]^{\mu+3/4}}e^{s_0/2},
\end{equation}
and
\begin{equation}
\lim_{\nu \rightarrow \infty} \ _1F_1\left(\mu-\nu+\frac{1}{2},2\mu+1,s_0\right)\approx
\frac{\Gamma(2\mu+1)}{\sqrt{\pi}}\frac{\cos[2\sqrt{s_0\nu}-\mu\pi-\pi/4]}{(s_0\nu)^{\mu+1/4}}e^{s_0/2}.
\end{equation}
By using these asymptotic forms for the confluent hypergeometric functions in Eq. \eqref{MITeq1}, the main
contribution reads
\begin{equation}
\frac{2\alpha \sqrt{m\omega}}{(\mathcal{E}+m)}\cos[2\sqrt{s_0\nu}-\mu\pi-3\pi/4]\approx 0,
\end{equation}
which is satisfied if
\begin{equation}
2\sqrt{s_0\nu}-\mu\pi-3\pi/4\approx(n+1/2)\pi,
\end{equation}
with $n$ being an integer number.
Now, by using \eqref{eq19} and noting that $s_0=m\omega r_0^2$, the energy levels are written as
\begin{equation}
\label{eq43}
\mathcal{E}_{n,l}\approx \pm\sqrt{m^2-m\omega\alpha(\alpha-2\kappa)
+\frac{\alpha^2\pi^2}{r_0^2}\left(n+\frac{\sqrt{\alpha^2+4\kappa(\kappa+\alpha)}}{4\alpha}+\frac{5}{4}\right)^2}.
\end{equation}
As expected, the energy spectrum of the fermionic field is modified by the confining potential.
The Eq. \eqref{eq43} represents the energy spectrum of the DO subject to the hard-wall potential in
the global monopole spacetime. We can see from the comparison between of Eqs. \eqref{eq21} and \eqref{eq43}
that the presence of the hard-wall potential modifies the energy levels of the DO. We can also observe that the
energy spectrum is influenced by the spacetime topology through the presence of the parameter associated with
the topological defect in the energy levels. We can note that by taking $\omega\rightarrow0$ into Eq. (\ref{eq43})
we obtain the energy spectrum of a Dirac field under effects of a hard-wall confining potential in the global monopole spacetime.
In addition, by making $\alpha \rightarrow 1$, we recover the energy
spectrum of the DO subject to the hard wall potential in Minkowisk spacetime.
We have plotted the behavior of the positive energy levels of this configuration in Fig. \ref{fig03}. In the left plot, the energy profile is shown
as function of $\omega/m$ for different values $\alpha$. In the right
plot, the energy profile is shown as function of the parameter  which characterises the presence of the topological defect.
Comparing these plots with the ones in the absence of the confining potential for the DO, Fig. \ref{fig01}, we note that the presence of the
confining potential increases the positive energy. But this energy still presents a decreasing behavior with the presence of the
global monopole. In addition, for small values of the $\omega/m$, the positive energy presents a contribution of the confining potential,
as expected.

\subsection{KGO}\label{sec4b}

In this subsection, we confine the scalar field subject to the KGO in the pointlike global monopole spacetime (\ref{eq01})
to a wall-rigid potential. First, we will consider the Dirichlet boundary condition, where the scalar field
vanishes at some fixed point $s_0$:
\begin{equation}
R(s_0)=0.
\label{dirichlet}
\end{equation}
Considering the radial wave function (\ref{eq34}), and the limit where $\bar{\nu}\gg 1$, and  proceeding in a similar
way done in the previous section, we find
\begin{eqnarray}
\label{eq45}
\mathcal{E}_{l,n}\approx\pm\sqrt{m^2+3m\omega\alpha^2+\frac{\alpha^2\pi^2}{r_0^2}\left(n+\frac{\sqrt{\alpha^2+8\xi(1-\alpha^2)+4l(l+1)}}{4\alpha}+\frac{3}{4}\right)^2},
\end{eqnarray}
which gives us the energy spectrum of a scalar field under effects of the KGO and to a hard-wall confining potential in the pointlike global monopole spacetime. We can note, by comparing Eqs. (\ref{eq33}) and (\ref{eq45}), that the presence of the hard-wall confining potential in the relativistic quantum system modifies the energy levels of the Klein-Gordon oscillator. We can also note that by taking $\omega\rightarrow0$ in Eq. (\ref{eq45}), we obtain the relativistic energy levels of a massive scalar field subject to the rigid-wall potential in the global monopole spacetime. In addition, for $\alpha\rightarrow1$ into Eq. (\ref{eq45}), we recover the energy spectrum of the KG oscillator in the Minkowski spacetime.

\begin{figure}[h]
	\centering
	{\includegraphics[width=0.48\textwidth]{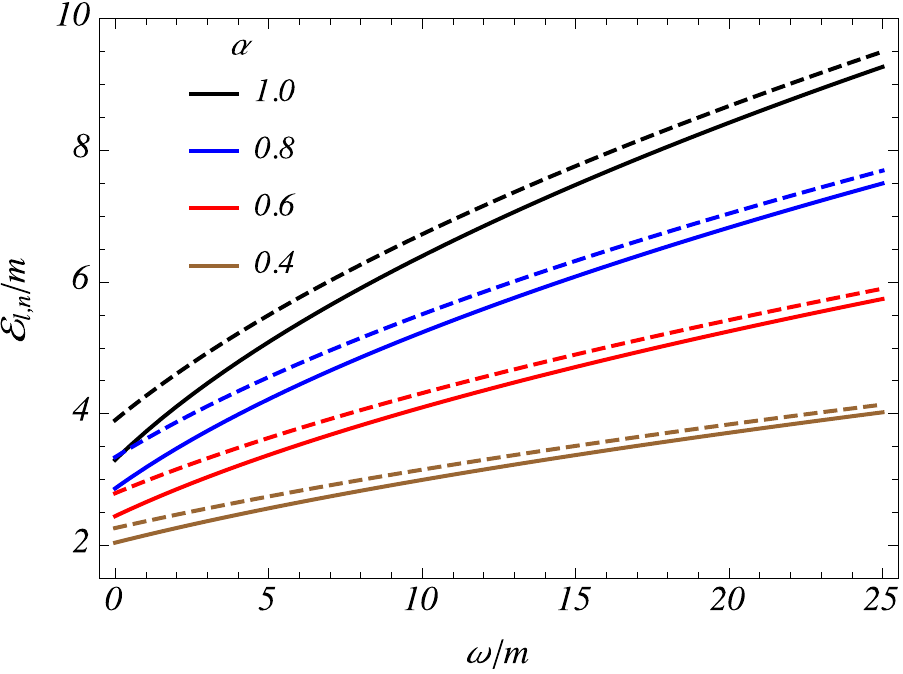}}
	\hfill
	{\includegraphics[width=0.48\textwidth]{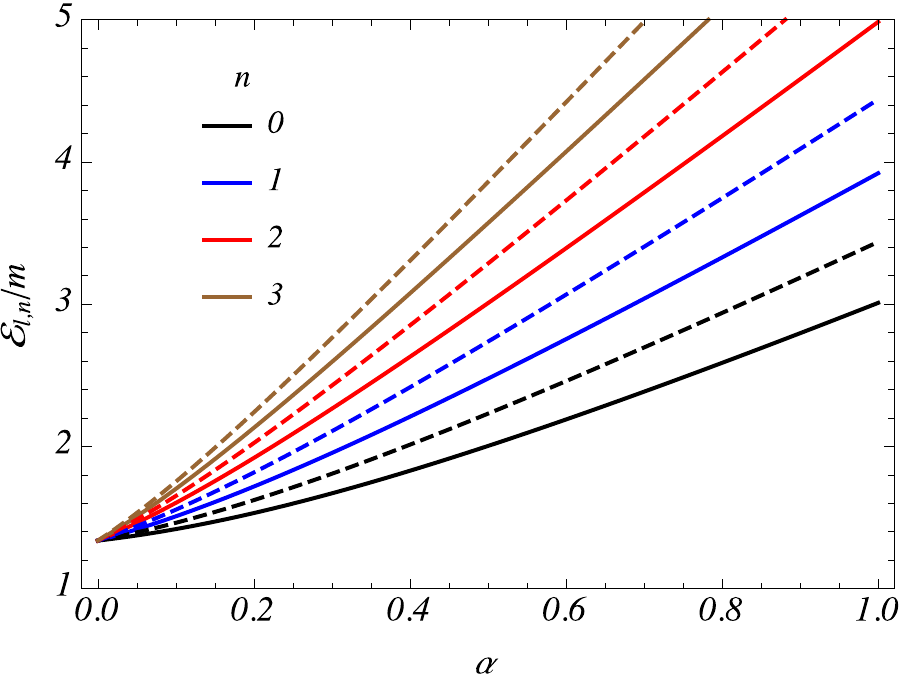}}
	\caption{Positive energy profiles of the KGO subject to a hard wall potential. In the left plot we have the energy as function of
	$\omega/m$ considering different values of the parameter $\alpha$. The right plot represents the energy as function of
	the parameter $\alpha$ for different energy levels and $\omega/m$. In both plots we have $mr_0=2.5$, $l=1$ and $\xi=0$. The full and
	dashed curves represents the Dirichlet and Neumann boundary conditions, respectively.}
	\label{fig04}
\end{figure}

We also can analyze the behavior of the energy levels considering the Neumann boundary condition:
\begin{equation}
\frac{d}{ds}R(s)|_{s=s_0}=0.
\label{neumann}
\end{equation}
By making use of the Eq. \eqref{eq34}, and following similar steps in comparison of the DO, the energy levels are given by
\begin{equation}
\mathcal{E}_{l,n}\approx\pm\sqrt{m^2+3m\omega\alpha^2
+\frac{\alpha^2\pi^2}{r_0^2}\left(n+\frac{\sqrt{\alpha^2+8\xi(1-\alpha^2)+4l(l+1)}}{4\alpha}+\frac{5}{4}\right)^2}.
\label{eq47}
\end{equation}
From the Eqs. \eqref{eq45} and \eqref{eq47} we note that the presence of a hard-wall confining potential modifies the relativistic energy levels of the system through the presence of the fixed radius $r_0$. In addition, the quantum numbers of the system have quadratic values, in contrast with Eq. (\ref{eq33}). The relativistic energy levels (\ref{eq45}) and \eqref{eq47} are influenced by the spacetime topology through the presence of the parameter associated with the topological defect $\alpha$. By making $\alpha\rightarrow1$, we recover the relativistic energy levels of a scalar field subject to the effects of the KGO and to a hard-wall confining potential in the Minkowski spacetime. The positive energy profiles \eqref{eq45}
and \eqref{eq47} are plotted in Fig. \ref{fig04}. In the left, plot the positive energy is shown as function of $\omega/m$ for
several values of $\alpha$ and a fixed energy level, while in the second one the positive is shown as function of
the parameter $\alpha$ for several energy levels, considering both Dirichlet (full curves) and Neumann (dashed curves) boundary conditions.
From both plots, similarly to the DO case, the presence of the confining potential increase the values of the energy. But this energy, still presents
a decreasing behavior with the presence of the global monopole. Besides, as we expect, when $\omega/m\rightarrow 0$ the positive energy
presents an extra contribution of the potential.

\section{Conclusions}
\label{sec5}
In this paper, we have investigated the influence of the global monopole in the DO and KGO relativistic quantum oscillators.
These oscillators are characterised by the changes: $\partial_r\rightarrow \partial_r \,+m\omega\beta r$
and $\hat{p}_\mu\rightarrow\hat{p}_\mu+im\omega X_\mu$, in the Dirac and the Klein-Gordon equations, respectively.
Taking the nonrelativistic limit, these modified equations are simplified to the Schr\"ondinger one.

We started our discussion analising the interaction between a spin-$\frac{1}{2}$ fermionic field and the DO. We solve the Dirac equation \eqref{eq10}
analitically, by using the complete set of functions \eqref{eq12} and we found the energy spectrum \eqref{eq21} and the radial solutions
\eqref{eq22}. Both, the energy spectrum and the radial solutions, are influenced by the presence of the parameter
which characterises the presence of the global monopole. The $|\mathcal{E}_{l,n}|$ is diminished when we compare with
the DO in the Minkowisk spacetime and as $\omega/m \rightarrow 0$, $\mathcal{E}_{l,n} \rightarrow \pm m$. These behaviors
are shown in Fig. \ref{fig01}, where we have plotted the profile of the energy of the DO as function
of $\omega/m$. In addition, this figure shows how the energy of the oscillator increases with the energy levels.
We also have plotted the influence of the topological defect
on the radial solutions in Fig. \ref{functionsDO}. In this figure, we have shown that in the regions near the defect, the radial solutions
is highly affected by it, while for the regions far away from the defect, its influence on the radial solutions becomes negligible.

As the next step, we have considered the interaction between a scalar scalar field and the KGO by solving analitically the
Klein-Gordon equation \eqref{eq23}. By considering the solution in the form \eqref{eq25}, we have found the energy spectrum
\eqref{eq33} and the radial eigenfunction \eqref{eq34}. Both, the energy spectrum and the radial solution, presents a influence of the global monopole.
In comparison with the KGO in the Minkowisk spacetime, the $|\mathcal{E}_{l,n}|$ are diminished and the energy tends to the
mass of the oscillator as $\omega/m$ goes to zero, as we can see in 
the Fig. \ref{fig03}, where we have plotted the
profile of the spectrum of the KGO.
In the Fig. \ref{functionKGO}, we have plotted how the defect influences the radial eigenfunctions \eqref{eq34}.
This influence is higher in the regions near the defect and becomes negligible in the regions
far away from the defect.

We have extended our discussion by investigating the influence of a hard-wall confining potential in our system.
For the DO, we have imposed that the fermionic field obeys the MIT-bag boundary condition \eqref{MIT}
at some fixed radius. For this configuration, we have found the energy spectrum \eqref{eq43} and plotted
the profile of the positive energy in the Fig. \ref{fig03}.
For the KGO we have considered the hard-wall confining potential by imposing that 
the scalar field obeys both the Dirichlet, \eqref{dirichlet}, and Neumann, \eqref{neumann}, boundary conditions at some fixed radius.
For these boundary conditions, the energy spectrum are given by the Eqs. \eqref{eq45} and \eqref{eq47}, respectively.
In the Fig \ref{fig04}, we have show the influence of the confining potential in the positive energy spectrum of the DO.
For both oscilattors, we have shown from Figs. \ref{fig03} and \ref{fig04}, that the confining potential increase the value of
the positive energy of the oscillators but still keeps its decreasing behavior in the presence
of the global monopole. In addition, as $\omega/m\rightarrow 0$ the positive energies of the DO and KGO presents a contribution of the
mass of the oscillator and another contribution of the potential, as we expect.

As mentioned in the introduction, our results can be applied to condensed matter systems. For example, the Refs. \cite{mg7b, tm3a} have investigated the harmonic oscillator in an environment with a point-type defect, such as vacancies and impurities, through the global monopole metric, which we based on investigating relativistic harmonic effects in this background. Besides that,
recently, thermodynamic properties of quantum systems have been investigated, for example, in diatomic molecule systems \cite{tm}, in a neutral particle system in the presence of topological defects in a magnetic cosmic string background \cite{tm1}, in exponential-type molecule potentials \cite{tm2}, on a 2D charged particle confined by a magnetic and Aharonov-Bohm flux fields under the radial scalar power potential \cite{tm3} and on a harmonic oscillator in an environment with a pontilike defect \cite{tm3a}. In particular, the DO and KGO oscillators have been the subject of research on thermodynamic properties, as shown in Refs. \cite{pt, pt1, pt2}. Therefore, the systems analyzed in this work can be a starting point for investigating the influence of the pointlike global monopole on the thermodynamic properties.

\acknowledgments{The authors would like to thank CNPq (Conselho Nacional de Desenvolvimento Cient\'ifico e Tecnol\'ogico - Brazil) and CAPES (Coordena\c{c}\~{a}o de Aperfei\c{c}oamento de Pessoal de N\'ivel Superior) for financial support. R. L. L. Vit\'oria was supported by the CNPq project No. 150538/2018-9.}

\end{document}